\documentclass[twocolumn,english,twocolumn,prd,epsf]{revtex4}
\usepackage[T1]{fontenc}
\usepackage[latin9]{inputenc}
\usepackage{amsmath}
\usepackage{color}
\usepackage{graphicx}

\makeatletter
\usepackage{amsfonts}
\usepackage{amsmath}
\usepackage{amssymb,amsbsy}
\usepackage{bm}
\usepackage{makeidx}
\usepackage{latexsym}
\usepackage{graphicx}
\usepackage{epsfig}
\usepackage{subfigure}
\usepackage{dsfont}
\usepackage{mathcomp}
\usepackage{color}
\usepackage{mathrsfs}

\makeatother

\usepackage{babel}

\begin{document}

\title{Effects of voids on the reconstruction of the equation of state of Dark Energy}

\author{Arnaud de Lavallaz\thanks{Electronic address: arnaud.de\_lavallaz@kcl.ac.uk}, 
Malcolm Fairbairn\thanks{Electronic address: malcolm.fairbairn@kcl.ac.uk}\\
\emph{Dept. of Physics, King's College London, Strand, London, WC2R 2LS, UK}}

\pagenumbering{arabic}
\newcommand{\ssection}[1]{\section[#1]{\centering\normalfont\scshape#1}}
\newcommand{\ssubsection}[1]{\subsection[#1]{\raggedright\normalfont\itshape#1}}

\begin{abstract}
We quantify the effects of the voids known to exist in the Universe upon the reconstruction of the dark energy equation of state $w$.  We show that the effect can start to be comparable with some of the other errors taken into account when analysing supernova data, depending strongly upon the low redshift cut-off used in the sample. For the supernova data alone, the error induced in the reconstruction of $w$ is much larger than the percent level.  When the Baryonic Acoustic Oscillations and the Cosmic Microwave Background data are included in the fit, the effect of the voids upon the determination of $w$ is much lessened, but is not much smaller than some of the other errors taken into consideration when performing such fits.  We also look at the effect of voids upon the estimation of the equation of state when we allow $w$ to vary over time and show that even when supernova, Cosmic Microwave Background and Baryonic Acoustic Oscillations data are used to constrain the equation of state, the best fit points in parameter space can change at the 10\% level due to the presence of voids, and error-bars increase significantly.
\end{abstract}

\maketitle

\section*{{\normalsize 1. INTRODUCTION\label{sec:1.-Introduction}}}
The current body of astronomical data makes it very difficult to live without dark energy.  Combination of the supernova data with the 7 year results from WMAP and the Sloan Digital Sky Survey (SDSS) data on Baryonic Acoustic Oscillations combine almost seamlessly with observations of the Hubble constant and the light element abundances to paint a picture which has become known as the standard cosmological model \cite{union2,wmap7,sdssbao,hstkey,bbn}.  The Universe is not homogeneous however and observations of galaxy clustering, N-body simulations and the simplest theoretical considerations all predict that the majority of the volume of today's Universe is considerably underdense in terms of matter, which forms roughly shaped denser walls surrounding the underdense voids \cite{2dfvoid,sdssvoid,milleniumvoid,Weygaert2009}.  The Universe is made up therefore mostly of the voids between overdense regions, a situation which has lead to evocative comparisons of the Universe with both a froth of bubbles and with Swiss Cheese \cite{swisscheese,swisscheese2}.

Such underdense regions have been studied in detail to understand their effects on cosmological observables \cite{romano1,romano2,tamara} and to investigate whether they can help explain the various observations of the Universe without dark energy \cite{Void1,Void2,Void3,Void4,Void5} - this might be possible if we are located close to the centre of a very large void \cite{Moffat1, Moffat2, Tomita1, Tomita2}.  Such a void would typically have to be larger than 100 Mpc in radius, in comparison to the voids which have definitely been observed in the Universe which vary in size between very roughly 5-20 Mpc.  In order to create such large voids, it would be necessary to have a very non-trivial power spectrum of perturbations \cite{voidpower1,voidpower2}. In this work we attempt to reach a much less ambitious goal and we only consider the voids which we have evidence to believe actually exist.

The nature of dark energy is unknown.  In particular, it is not known if its energy density remains completely constant over time or whether it is growing or decreasing.  If dark energy can be modelled as a perfect fluid then any such evolution is determined by the equation of state of the fluid $w$ which relates the energy density and the pressure, $P=w\rho$.  By fitting supernova data, we are able to constrain that equation of state and future investigations hope to pin it down with increasing accuracy.  

Voids can be considered not only to provide additional constraints on $w$ \cite{w1,w2}, but also as potential sources of error. In this study we estimate what the effect of voids will be upon the reconstruction of the equation of state of dark energy, $w$, from the supernova data.  In order to do this, we look at the evolution of voids in a Universe that contains collisionless matter and a cosmological constant.  We consider distant voids in which supernovae explode and look at the effect upon photons as they climb out of their host void walls and move towards us.  The presence of the voids increases the scatter in the luminosity redshift plane - we aim to estimate this error and see how it compares with other errors on the reconstruction of $w$.  We do not expect the error to be large since the fluctuations in the gravitational potential will be very small, less than a percent.  However, since the velocity dispersion of supernovae relative to the Hubble flow can create an observable effect, we feel it is worthwhile investigating the magnitude of the error due to these smaller voids.

We do this in a way which is not entirely self-consistent - this shortcoming lies in the fact that we evolve voids in a $\Lambda$CDM Universe and then apply the estimated error to other kinds of dark energy where $w\neq-1$.  The reason we do this is that although it is in principle possible to study voids evolving in a background of dark energy with $w\neq -1$, the evolution of such voids involves the solution of a complex set of partial differential equations, whereas voids evolving in a cosmological constant background are the result of ordinary differential equations and are therefore simpler to analyse numerically.  We hope to estimate the magnitude of the error one expects from these walls in the knowledge that when one includes Baryonic Acoustic Oscillations (BAO) data and the Cosmic Microwave Background (CMB), the equation of state $w$ is constrained to lie quite close to $-1$ anyway.  We note that much more complicated analyses might be done in future if it becomes necessary.

In the next section we will outline the equations that we evolve as well as the functional form of the void models we choose.  We will then describe our numerical procedures and present our results before making conclusions.

\section*{{\normalsize 2. THE LTB EQUATIONS AND VOIDS\label{sec:1.-Analytical_Considerations}}}

Although we will assume that all the matter in the Universe has evolved into voids and void walls, our interest is focused on the hole nearest to the source only.   We assume that the effects of \emph{complete} holes situated between the source and the observer cancel out along the light path and we neglect the void that we are located in since all photons must arrive into the same current potential well.  We construct our Swiss-Cheese model of the universe by combining two components: the 'cheese', where the metric is described by a spatially flat Friedmann-Robertson-Walker (FRW) solution and the 'holes', where the matter is mainly underdense by volume and we use the Lema\^itre-Tolman-Bondi solution.  Unlike the idea that we are at the centre of a very large void which is used to live without dark energy, we will therefore be in the cheese, looking at distant holes.

\subsection*{A. The Model: Lema\^{i}tre-Tolman-Bondi Universe}

Our model is based on the Lema\^{i}tre-Tolman-Bondi (LTB) metric,
a spherically symmetric solution of Einstein's equations, which can
be written as (in units where $c=1$)\begin{equation}
ds^{2}=-dt^{2}+S^{2}(r,t)dr^{2}+R^{2}(r,t)(d\theta^{2}+sin^{2}\theta d\phi^{2})\label{eq:LTB0}\end{equation}
where we use comoving coordinates $(r,\theta,\phi)$ and proper time
$t$. For pressureless dust and a cosmological constant, Einstein's equations imply the
following constraints:\begin{equation}
S^{2}(r,t)=\frac{R^{\prime2}(r,t)}{1+2E(r)},\label{eq:LTB1}\end{equation}
\begin{equation}
\frac{1}{2}\dot{R}^{2}-\frac{GM(r)}{R(r,t)}-\frac{1}{3}\Lambda R^{2}=E(r),\label{eq:LTB2}\end{equation}
\begin{equation}
4\pi\rho(r,t)=\frac{M^{\prime}(r)}{R^{\prime}(r,t)R^{2}(r,t)},\label{eq:LTB3}\end{equation}
where a dot stands for partial derivative with respect to $t$ and
a prime with respect to $r$; $\rho(r,t)$ is the energy density of
the matter, $G$ is Newton's constant and $\Lambda$ is the cosmological constant. To
specify the model we intend to use, we have to define the two arbitary
functions $E(r)$, corresponding to the spatial curvature, and $M(r)$,
which is simply the mass integrated within a comoving radial coordinate
$r$:\[
M(r)=4\pi\int_{0}^{r}\rho(r,t)R^{2}R^{\prime}dr.\]
Both of these functions are completely defined therefore by the choice of an initial density profile $\rho(r,t_{\text{LTB}})$,
where $t_{\text{LTB}}=t_{\text{LTB}}(r)$ refers to the beginning
of the LTB evolution and is set to a constant for simplicity. We then
have to choose $E(r)$ in order to match the flat FRW model at the
boundary of the hole. This choice also ensures
that the average density inside the hole equals the one outside, so
that an observer situated in the cheese would not be aware, \textit{locally},
of the presence of the hole.

We intend to test our model by sending photons through the voids and
measuring redshifts. To do so, we must integrate the equation for
their radial trajectory\begin{equation}
\frac{dt(r)}{dr}=\frac{R^{\prime}\left(r,t(r)\right)}{\left(1+2E(r)\right)^{1/2}}.\end{equation}

To repeat and clarify, in contrast to other models based on the Swiss-Cheese universe, we assume that the observer, although occupying no particular position in space with respect to the holes, is always situated in the cheese; we then consider only the effects caused by the furthest hole where the supernovae are located (see Subsection 3.A for more details) and hence the geodesics originate since in this study we want to investigate the effect of the void at the origin of the photons rather than when they arrive here.

\subsection*{B. Parameters and Initial Conditions}

\begin{figure}
\includegraphics[scale=0.48,angle=-90]{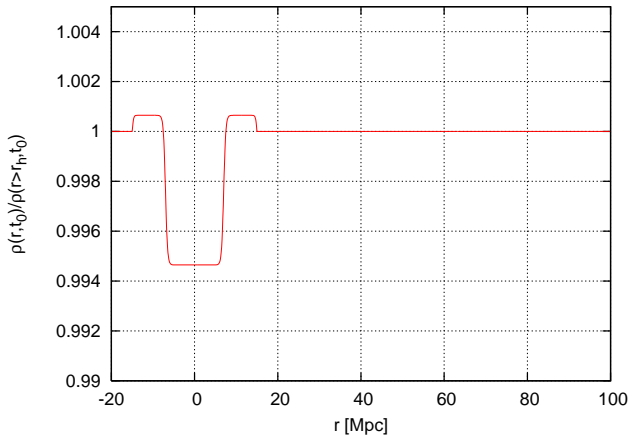}

\caption{Initial density profile define at $t=t_{\text{LTB}}$ and showing
the shape of the cosmological voids and of its surrounding walls of
denser matter.\label{fig:Initial-density-profile}}

\end{figure}

To characterise the LTB model we intend to use, we have to
define the initial density profile $\rho(r,t_{\text{LTB}})$. We then
will build the two arbitrary functions $E(r)$ and $M(r)$ based on
this profile. The expression we choose for it is based on Kostov's parametrisation \cite{Kostov2009} where the density profile (Fig. \ref{fig:Initial-density-profile}) is defined as follows:\begin{multline}
\rho(r,t_{0})=\bar{\rho}(t_{0})\times\\
\left\{ A_{1}+A_{2}tanh\left[\alpha\left(r-r_{1}\right)\right]-A_{3}tanh\left[\beta\left(r-r_{2}\right)\right]\right\} ,\end{multline}
for $r<r_{h}$, where $r_{h}$ is the radius of the void; and $\rho(r,t_{0})=\bar{\rho}(t_{0})$
for $r\ge r_{h}$. Given $(r_{1},r_{2})$, the values of the coefficients
$(A_{1},A_{2},A_{3})$ and $(\alpha,\beta)$ are chosen so that $\rho(r_{h},t_{0})=\bar{\rho}(t_{0})$
and the integrated mass inside the hole $M(r_{h})=4\pi\int_{0}^{r_{h}}\rho(r,t_{0})r^{2}dr=\frac{4}{3}\pi\bar{\rho}(t_{0})r_{h}^{3}$.

The function $E(r)$ plays the role of the spatial curvature in the
LTB solution. It is defined using Eq. \ref{eq:LTB2} and considering
the initial time of the LTB evolution $t_{\text{LTB}}$. At this early
time, $R(r,t_{\text{LTB}})=a(t_{\text{LTB}})r=a_{\text{LTB}}r$ and
we have, in real units:\begin{equation}
E(r)=\frac{1}{2}\frac{H_{\text{LTB}}^{2}a_{\text{LTB}}^{2}}{c^{2}}\left(r^{2}-\frac{3}{4\pi}\frac{M(r)}{a_{\text{LTB}}^{3}r\bar{\rho}(t_{\text{LTB}})}\right),\end{equation}
where $\bar{\rho}(t)$ is the average energy density at time $t$.

In order to guarantee that the chosen parametrisation of the voids
is consistent with observational constraints, we compare the density
profiles with observed values of over- and underdensities at different
times of the LTB evolution: at initial time $t=t_{\text{LTB}}$ and
at final time $t=t_{\text{now}}$. We set $t_{\text{LTB}}$ to be
the time of last scattering, $t_{\text{LTB}}=t(z=1100)$. So the constraint
at initial time comes from the scale of temperature fluctuation in
the Cosmic Microwave Background (CMB) $\Delta T/T\approx10^{-5}$,
which corresponds to a variation in density of the order of $10^{-4}$.

At the other extremity, when $t=t_{\text{now}}$, the guideline for
density profiles is given by observations of the present underdensities
in the matter distribution of our universe. Recent studies (e.g. \cite{Weygaert2009})
show that the voids that we see nowadays probably correspond to regions
where the density is $\sim20\%$ of the mean cosmic density.

Further constraints on the size of the voids can be deduced from the 
study of non-linear late-time integrated Sachs-Wolfe effects on the 
CMB \cite{Valkenburg}.

Ensuring that these constraints are taken into account defines the spatial geometry of the voids. The chosen
parameters are given in Table \ref{tab:Parameters-Voids}.%
\begin{table}
\begin{tabular}{|c|c|} 
\hline 
\multicolumn{2}{|c|}{Density Profile Parameters} \\ 
\hline 
$r_1$ & 4 Mpc \\ 
$r_2$ & 15 Mpc \\
\hline 
$A_1$ & $9.97\times 10^{-1}$ \\
$A_2$ & $2.99\times 10^{-3}$ \\
$A_3$ & $1.70\times 10^{-4}$ \\
\hline 
$\alpha$ & 0.6 \\
$\beta$ & 3.0 \\
\hline
\end{tabular}
\caption{Chosen parametrisation of the voids, taking into account the constraints
at initial time $t_{\text{LTB}}$ and final time $t_{\text{now}}$.\label{tab:Parameters-Voids}}
\end{table}

\section*{{\normalsize 3. SIMULATION PROCESS AND RESULTS\label{sec:1.-Results}}}

\subsection*{A. Obtaining Redshifts and Distance Moduli\label{sub:zdL}}
In order to compare the various Universes with dark energy possessing different equations of state,
we need to build Hubble diagrams which allow us to analyse the discrepancies
between models and Type Ia Supernova data. We therefore
 obtain the redshift of distant sources situated around the void
and then calculate the corresponding distance modulus.

We evolve a single void up to some random time and then choose a random position in the void according to the weighting of matter density.  We assume that the number density of supernovae explosions is proportional to the density and therefore more likely to occur in the void wall. We follow the path of a photon emerging from that void, with initial coordinates $(r_{in},t_{in})$, until the value of $t(r)$ reaches the present time, $t(r_{\text{final}})=t_{\text{now}}$.  
We then send a second photon, with initial coordinates $(r^{\prime}_{in}=r_{in},t^{\prime}_{in}=t_{in}+\Delta t_{in})$, until the comoving distance it covered corresponds to $r_{\text{final}}$. At this point, we can calculate $\Delta t_{\text{final}}=t^{\prime}(r_{\text{final}})-t_{\text{now}}$ and thus the redshift $z(r_{\text{in}},t_{\text{in}})=\Delta t_{\text{final}}/\Delta t_{\text{in}}-1$.  
The luminosity distance is then given by 
\[d_{L}=(1+z)^2 R(r_{\text{final}},t_{\text{now}})=a_0r_{\text{obs}}(1+z),\]
where $a_{0}$ is the value of the scale factor today and $r_{\text{obs}}=r_{\text{fin}}-r_{\text{in}}$ is the distance between the source and the observer. We repeat this process, generating a cloud on the distance modulus / redshift plane which takes the form of a blurry line with scatter around the normal $\mu-z$ relationship in $\Lambda$CDM.

\subsection*{B. Statistics: Binning Process\label{sub:Binning_Process}}

\begin{figure}
\begin{centering}
\includegraphics[scale=0.45,angle=-90]{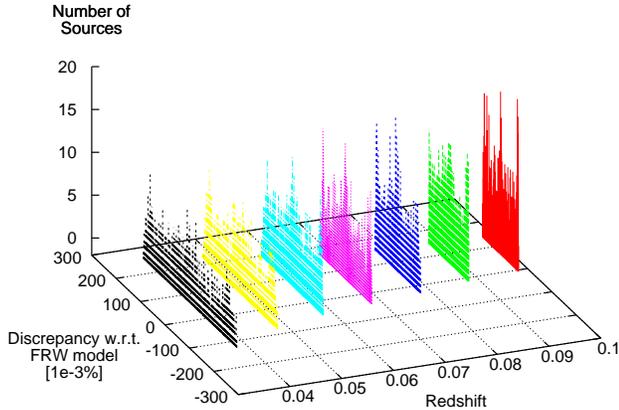}
\par\end{centering}

\caption{Illustration of the binning process allowing us to deal with the statistical
nature of our results. The width of each bin is fixed in redshift,
$\Delta z=0.01$.\label{fig:Binning}}

\end{figure}

The aim of the binning process is to gather series of points given
by simulations in order to deduct sensible estimates of the error on the redshift and luminosity distance from them.
Given the redshift range covered by the supernova data, $0.01<z<2$, we choose to
distribute the 10$^{5}$ data points between 200 bins of width $\Delta z=0.01$ (Fig. \ref{fig:Binning}).
For each bin obtained, the mean and standard deviation are calculated.
We thus end up with a new series of points and corresponding errorbars
which we can now add to the existing known errorbars on the supernovae and see how the cosmological conclusions change.

The Type Ia Supernovae data used in this study are extracted from
the Union2 Compilation \cite{union2} of the Supernova Cosmology
Project, which contains 557 sources drawn from 17 datasets. We base
our comparison on the file called ``Union2 Compilation Magnitude
vs. Redshift Table'' \footnote{http://www.supernova.lbl.gov/Union}, 
which contains supernova name,
redshift, distance modulus, and distance modulus error. To
the latter we add the systematic error due to the peculiar velocities
of the host galaxies, $\sigma_{v}=400$ km/s, converted into a systematic
error on the redshift of $\delta z=0.00132$.

Once all data are binned and all errors known, we can compute $\chi^{2}$
values for every combination $\left(\Omega_{m},\Omega_{\Lambda}\right)$.
In parallel, we do the same for all FRW universes defined by the same
density parameters.

\subsection*{C. Comparison of Errors}

To compute consistent $\chi^{2}$ values, we ought to take
into account all the errors and uncertainties adding up for each source
considered and which can be split into two categories: error on distance
modulus, $\sigma_{\mu}$, and binning process uncertainty, $\sigma_{\text{BP}}$,
so that\[
\chi^{2}=\sum_{\text{SNe}}\frac{\left[\mu_{\text{mean}}(z)-\mu_{\text{meas}}(z)\right]^2}{\sigma_{\mu}^{2}+\sigma_{\text{BP}}^{2}}.\]

\textsc{\small Error on Distance Modulus |} This first error gathers
the variety of measurement errors due to the nature of the sources
and the way their luminosity is evaluated. Accordingly to the Union2
original paper \cite{union2}, we sum in the component errors
coming from the following causes: light curve fitting (through covariant
matrix), galaxy peculiar velocities, Galactic extinction, gravitational
lensing and a floating dispersion term (for potential sample-dependent
systematic errors).\\
All these errors are included in the error on distance modulus given
by the Union2 Compilation data.

\medskip{}

\textsc{\small Binning Process Uncertainty |} The second group of errors relate to the spread induced by the voids themselves. 

Here the error comes from the binning process itself, where
we combine results of simulation for a large number of different sets
of initial conditions. For each bin created at a given redshift $z$
(see Subsection 3.B), we determine the standard deviation of the data
by evaluating the spans from the mean value that include 68\% of all
the points in the bin. The corresponding deviation, $\sigma_{\text{BP}}(z)$,
is then simply added in quadrature to the total uncertainty. 

\subsection*{D. Effect on the estimates of the Equation of State $w$}

Having estimated the errors due to the growth of voids in the universe, we are in a position to find out how this effects the reconstruction of the equation of state.  To do this we perform a Monte Carlo Markov Chain (MCMC) fitting to the data.  We assume a flat Universe and generate values of $\Omega_M$ (and therefore $\Omega_{DE}=1-\Omega_M$) and $w$ randomly. For all fits in this paper we remove uncertainty in the Hubble constant by choosing the best fit value for each value of $\Omega_M$. We also calculate the probability from the $\chi^2$ values using the $\chi^2$ cumulative distribution function rather than relative $\chi^2$ values as an estimation of maximum likelihood.

\begin{figure}
\begin{centering}
\includegraphics[scale=0.82]{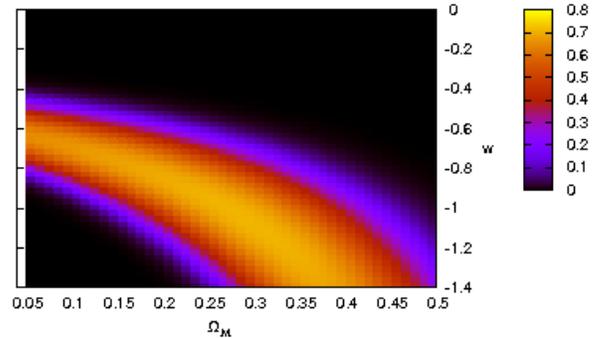}
\par\end{centering}

\caption{$\Omega_M$ vs dark energy equation of state $w$ when fitting the Union2 data set only assuming a flat Universe.  The shading represents probability from the $\chi^2$ cumulative distribution function. \label{snonlyshaded}}

\end{figure}

First we fit the supernova data alone.  Typical favoured regions in the $\Omega_M-w$ plane can be seen in figure \ref{snonlyshaded}.  Table \ref{snonly} shows the projected constraints on the equation of state with and without the presence of the additional error due to voids.  We have done MCMC runs for the supernova data sets with various minimum redshifts.  When fitting to data, it is normal to cut out very small redshift supernova since they are the data points most vulnerable to large redshift errors such as peculiar velocities.  Given that we expect the void effect to be at most the same magnitude as these other errors, we perform fits to the data with and without the voids neglecting supernovae below different minimum redshifts - $z_{min}=0.01, 0.02, 0.03$ and $0.04$.  As expected, the discrepancy between the simulations with and without voids decreases as we increase the minimum redshift.

The constraint on the value of $w$ when only including the supernova data is very weak and best fit values are quite far from $w=-1$.  This regime is therefore far from the region of validity of our analysis - the growth of voids in such a Universe might be quite different to the LTB case due to structure formation in the dark energy and anisotropic pressure which in general is a challenge to calculate.  However, we make no apologies for this as we only seek to highlight the possible magnitude of void effects upon the reconstruction of the equation of state in this work.

\begin{table}
\begin{center}
\begin{tabular}{| l | c | c | c | }
\hline
$z_{min}$ & Smooth & With Voids & $|\Delta w|$\\
\hline
0.01 & $-1.480^{+0.631}_{-1.005}$ &$-1.613^{+0.740}_{-1.189}$ & 0.133 \\
0.02  & $-1.574^{+0.706}_{-1.125}$ &$-1.662^{+0.776}_{-1.250}$ & 0.088 \\
0.03  & $-1.727^{+0.819}_{-1.320}$ & $-1.823^{+0.889}_{-1.422}$ & 0.096 \\
0.04  & $-2.017^{+1.006}_{-1.678}$   & $-2.056^{+1.028}_{-1.748}$ & 0.039  \\
\hline
\end{tabular}
\caption{The best fit for the equation of state of dark energy $w$ to the Union2 supernova data set alone with the 67\% errors.  The fits are done for different minimum redshifts and with and without the additional errors created by the presence of voids.  The fourth Column is the difference between the two best fit values.  Note the general trend of the difference in errors going down as we cut out lower redshift supernovae (apart from when we go from $z_{min}=0.02$ to $z_{min}=0.03$ which must be a statistical fluctuation).}
\label{snonly}
\end{center}
\end{table}

\begin{figure}
\begin{centering}
\includegraphics[scale=0.82]{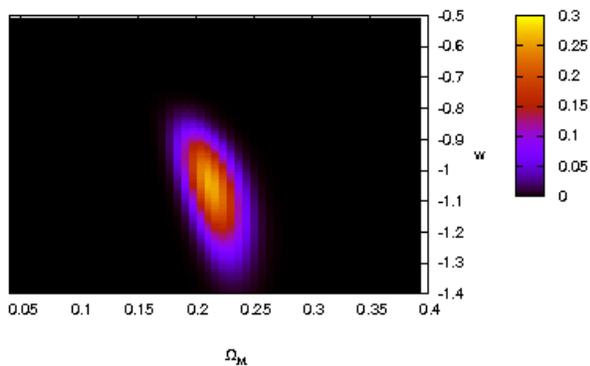}
\par\end{centering}
\caption{$\Omega_M$ vs dark energy equation of state $w$ when fitting the Union2 data set, Baryonic Acoustic Oscillations and the shift parameter and acoustic scale of the CMB, again assuming a flat Universe.  The shading represents probability from the $\chi^2$ cumulative distribution function. \label{snbc}}
\end{figure}

Next we include the effective distance measure $D_{V}$ which is obtained from the SDSS Baryonic Acoustic Oscillations data \cite{eisentein2005,sdssbao} and the WMAP constraint on dark energy from the shift parameter $R$ and the acoustic scale $l_A$ \cite{wmap7}.  The favoured regions on the $\Omega-w$ plane can be seen in figure \ref{snbc}.  These lead to much tighter constraints on possible values of $w$ and the difference between the cases with and without voids is much smaller (Table \ref{sncmbbao}).

\begin{table}
\begin{center}
\begin{tabular}{| l | c | c | c | }
\hline
$z_{min}$ & Smooth & With Voids & $|\Delta w|$\\
\hline
0.01 & $-1.064^{+0.116}_{-0.126}$  & $-1.072^{+0.127}_{-0.139}$ & 0.008 \\
0.02 & $-1.073^{+0.122}_{-0.132}$  & $-1.078^{+0.129}_{-0.141}$ & 0.005 \\ 
0.03 & $-1.097^{+0.131}_{-0.143}$  & $-1.101^{+0.134}_{-0.147}$ & 0.004 \\
0.04 & $-1.128^{+0.134}_{-0.148}$  & $-1.129^{+0.135}_{-0.149}$ & 0.001 \\
\hline
\end{tabular}
\caption{The best fit for the equation of state of dark energy $w$ to the Union2 supernova data set, CMB and BAO, as described in the text, with the 67\% errors.  The fits are done for different minimum redshifts and with and without the additional errors created by the presence of voids.}
\label{sncmbbao}
\end{center}
\end{table}

We note here that we obtain slightly different constraints on the equation of state to the analyses carried out in \cite{union2}, which is probably mostly due to the way we marginalise over the Hubble constant. The basic preferred regions are compatible in the two studies.

%DISCUSSION ON DYNAMIC DARK ENERGY***********************

\subsection*{E. Effect on the estimates of the dynamic Equation of State $w(z)$}

So far, we have considered a constant value for the equation of state parameter $w$. It is however possible that the dark energy equation of state might vary over time and therefore redshift.  A very commonly used parametrization of dynamic dark energy is \begin{equation*} w = w_0 + w_a\frac{z}{1+z}, \end{equation*} based on works by Chevallier, Polarski and Linder \cite{dynamic1,dynamic2}. Following the same procedure as before, we fit the supernovae data while taking into account the constraints given by the CMB and the BAO. The results for the best fits are given in Table \ref{dynamicw}.

\begin{table}
\begin{center}
\begin{tabular}{| l | c | c | c| c | }
\hline
$z_{min}$ & $w_i$ & Smooth & With Voids & $|\Delta w_i|$\\
\hline
0.01 & $w_0$ & $-0.523^{+0.457}_{-0.320}$  & $-0.492^{+0.662}_{-0.636}$ & 0.031 \\
     & $w_a$ & $-2.993^{+6.956}_{-2.672}$  & $-3.223^{+8.166}_{-2.938}$ & 0.230 \\
0.02 & $w_0$ & $-0.515^{+0.587}_{-0.343}$  & $-0.424^{+0.426}_{-0.402}$ & 0.091 \\
     & $w_a$ & $-3.064^{+5.037}_{-2.572}$  & $-3.660^{+7.097}_{-3.096}$ & 0.596 \\
0.03 & $w_0$ & $-0.444^{+0.395}_{-0.394}$  & $-0.475^{+0.558}_{-0.417}$ & 0.031 \\
     & $w_a$ & $-3.488^{+6.608}_{-2.686}$  & $-3.272^{+5.751}_{-3.025}$ & 0.216 \\
0.04 & $w_0$ & $-0.483^{+0.953}_{-0.412}$  & $-0.493^{+0.489}_{-0.427}$ & 0.010 \\
     & $w_a$ & $-3.173^{+5.173}_{-2.756}$  & $-3.076^{+5.451}_{-2.798}$ & 0.097 \\
\hline
\end{tabular}
\caption{The best fit for the dynamic equation of state of dark energy $w = w_0 + w_a\frac{z}{1+z}$ to the Union2 supernova data set, CMB and BAO, with the 67\% errors.  The fits are done for different minimum redshifts and with and without the additional errors created by the presence of voids.}
\label{dynamicw}
\end{center}
\end{table}

\begin{figure}
\begin{centering}
\includegraphics[scale=0.9]{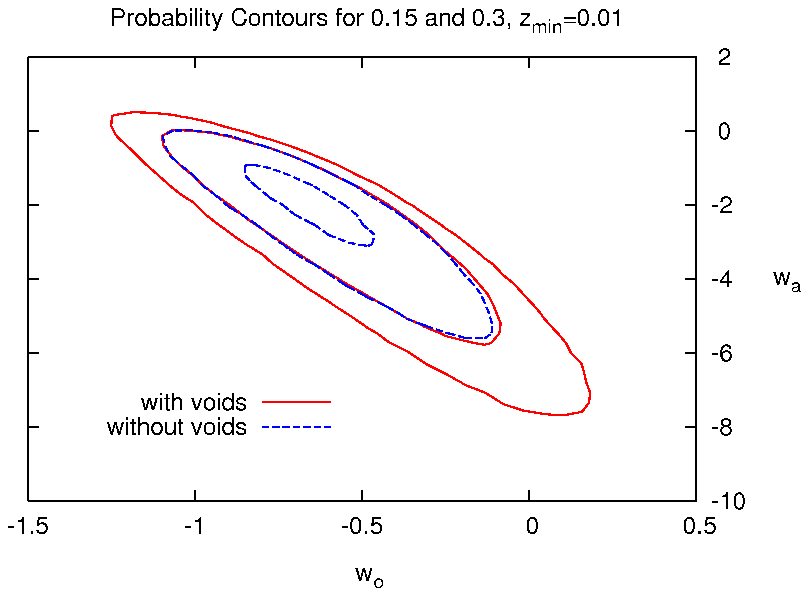}
\includegraphics[scale=0.9]{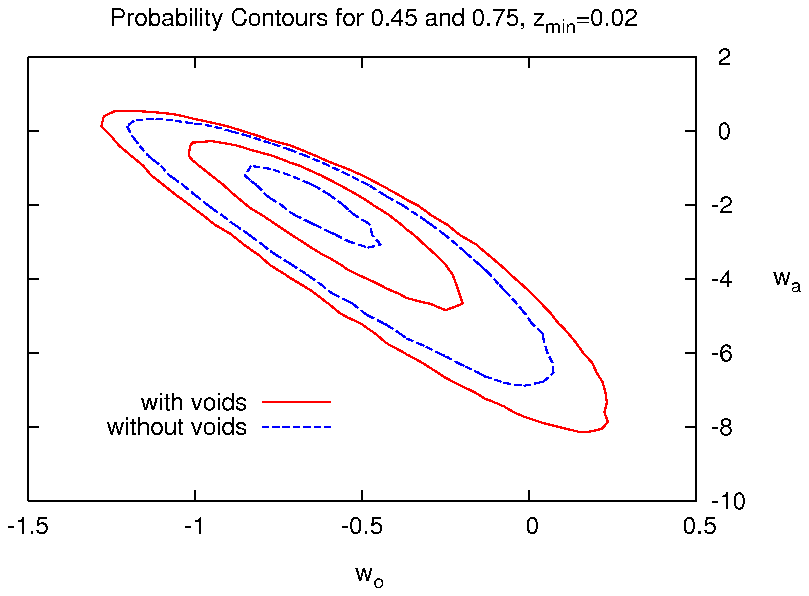}
\includegraphics[scale=0.9]{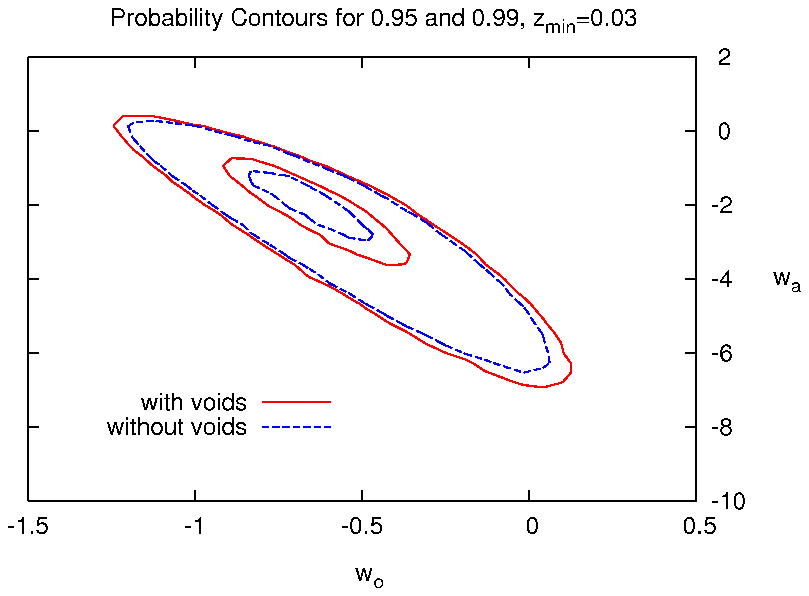}
\includegraphics[scale=0.9]{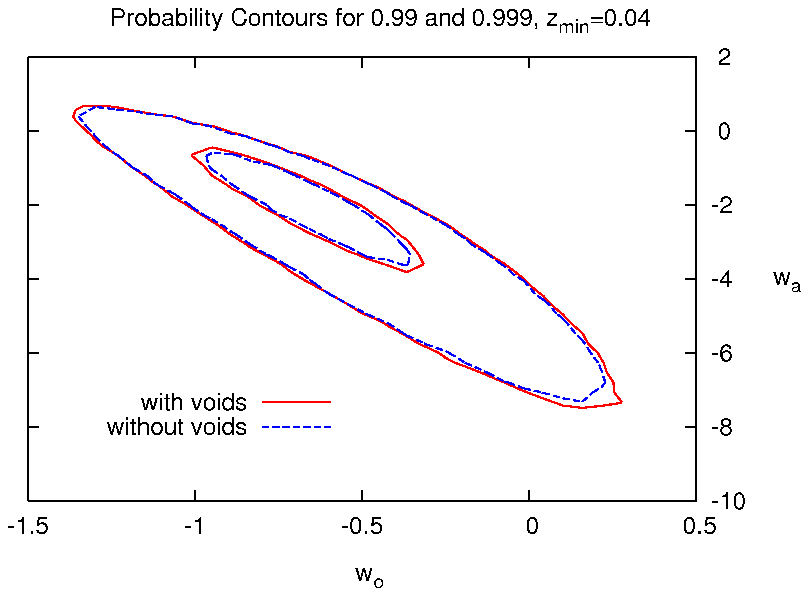}
\par\end{centering}
\caption{Dark energy equation of state parameters $w_0$ versus $w_a$, with $z_{min}=0.01,0.02,0.03\text{ and }0.04$ respectively, when fitting the Union2 data set, Baryonic Acoustic Oscillations and the shift parameter and acoustic scale of the CMB, always assuming a flat Universe.  The contours correspond to the probabilities listed in the individual graph titles.} 
\label{fig:dynamicw}
\end{figure}

Even with the extra degree of freedom, the fits obtained when including the whole set of supernovae down to arbitrai;y small redshifts are not fantastic, the probability obtained from the $\chi^2$ distribution of the best fit point in parameter space being less than 50\%. However, the situation changes dramatically depending on the redshift considered for the cut-off: the best fit probability comes close to 1 with $z_{min}=0.04$. Fig. \ref{fig:dynamicw} shows the contour plots obtained for the different values of $z_{min}$ studied here.

It is clear again that if we include the lowest redshift supernovae, the error which is introduced due to the presence of voids has a considerbale impact on the best-fitting regions in $w_o-w_a$ parameter space.

%END OF DISCUSSION ON DYNAMIC DARK ENERGY*****************

\section*{{\normalsize 4. COMMENTS AND CONCLUSIONS\label{sec:1.-Conclusion}}}

There are more than 100 supernovae below $z=0.04$ in the Union2 data set so it is not surprising that including or not including them can have a considerable effect on the reconstruction of the equation of state of dark energy $w$.  In the same way, changing their error bars changes their pull on the best fit values of $w$.

In this work, we have shown that the growth of voids in the Universe can have a small effect upon the reconstruction of the equation of state of dark energy by inducing scatter into the redshift-luminosity data plane.  If one considers only the supernova data, and we try to constrain models of dark energy with a constant equation of state, the effect of the voids may be relatively large, up to around 10\% of the central value of $w$, becoming less as one increases the minimum redshift supernovae included in the fit. 

When one includes the constraints from the CMB and from Baryonic Acoustic Oscillations, the magnitude of the effect is much smaller, below the percentage level. However, the induced error does start to compare with some of the other smaller errors that are considered by the Union2 team which are listed in table 9 of \cite{union2}.  This error is not important now and may turn out to be dwarfed by other systematics even in future studies using data from new observational programs.  Conversely, one can envisage a situation where it would have an important effect upon the understanding of dark energy and we would need to parametrise and quantify our uncertainty.

If we consider dark energy with an equation of state which is allowed to vary, i.e. $w(z)=w_o+w_az/(1+z)$, we find that the effect of the uncertainty introduced due to the voids is amplified and the best fit values of $w_o$ and $w_a$ can change around 10\%, unless we neglect all supernovae at redshifts lower than $z\sim 0.04$.  The magnitude of the errors around these best fit regions also increase significantly.

Another approach to understand the distribution of these voids and their effect is to try and map out the local voids using peculiar velocity flows in an attempt to quantify their effect.  This seems difficult : a redshift of $z\sim 0.04$ corresponds to a comoving distance of roughly 150 Mpc.  The size of typical voids in the Universe today is considerably smaller than this, of the order of 10-20 Mpc in radius.  We should therefore expect many hundreds of voids to be present in this low redshift part of the Hubble diagram and it seems impossible to trace them given our lack of understanding of bias and the history of structure formation in the local Universe.  What we do know is that there are voids, their gravitational potential may have implications for the reconstruction of $w$ and we can in principle quantify this uncertainty.

\section*{Acknowledgments}
MF would like to thank Rahman Amanullah for an interesting discussion and Nigel Arnot for repeatedly re-booting my server.  
The authors are grateful to Wessel Valkenburg for his advice and helpful comments.
AdL receives the benefit of funding from the KCL Graduate School.

\newpage{}

\end{document}